\documentclass{article}
\usepackage{microtype}
\usepackage{graphicx}
\usepackage{subfigure}
\usepackage{booktabs}
\usepackage{hyperref}

\usepackage{mlsys2025}
\def\isaccepted{1}
\makeatletter
\renewcommand{\Notice@String}{}
\makeatother
\newlength{\figcaptionskip}
\newlength{\tabcaptionskip}
\makeatletter
\long\def\@makecaption#1#2{%
  \def\@tempa{table}%
  \ifx\@captype\@tempa\else\vskip\figcaptionskip\fi
  \baselineskip 11pt
  \setbox\@tempboxa\hbox{#1. #2}%
  \ifdim \wd\@tempboxa >\hsize
    \sbox{\newcaptionbox}{\small\sl #1.~}%
    \newcaptionboxwid=\wd\newcaptionbox
    \usebox\newcaptionbox {\footnotesize #2}%
  \else
    \centerline{{\small\sl #1.} {\small #2}}%
  \fi
  \ifx\@captype\@tempa\par\vskip\tabcaptionskip\fi}
\makeatother
\usepackage{url}

\usepackage{amsmath,amssymb} %
\usepackage{color}
\usepackage{multirow}
\usepackage{comment}
\usepackage{stackengine}
\usepackage{colortbl}
\usepackage{amsthm}     %
\usepackage{wrapfig}
\usepackage{physics}
\usepackage{mathtools}
\usepackage{cleveref}
\usepackage{listings}
\usepackage{soul}
\usepackage{xspace}
\usepackage{xr}

\newcommand{\mainref}[1]{\ref{#1}}
\definecolor{gray}{gray}{0.5}
\newcommand{\sys}{\textsc{CompileMask}\xspace}  %
\soulregister\sys{0}

\lstdefinestyle{signature}{
    language=Python,
    basicstyle=\ttfamily\small\bfseries,
    keywordstyle=\color{magenta},
    morekeywords={int,bool},
    numbers=none,
    frame=none,
    breaklines=true,
    breakindent=1.5em,
    aboveskip=4pt,
    belowskip=4pt
}

\newtheorem*{proposition*}{Proposition}

\renewcommand{\algorithmicrequire}{ \textbf{Input:}} %
\renewcommand{\algorithmicensure}{ \textbf{Output:}} %

\newtheorem{theorem}{Theorem}
\newtheorem*{theorem*}{Theorem}



\sethlcolor{yellow}

\makeatletter
\DeclareRobustCommand{\aipolished}{\leavevmode\bgroup\color{blue}\let\aipolished@brace= }
\makeatother
\ifdefined{\def\aipolished{}}\fi

\usepackage[normalem]{ulem}
\newcommand{\aideleted}[1]{{\color{gray}\let\aidelcite\cite\renewcommand{\cite}[1]{\mbox{\aidelcite{##1}}}\sout{#1}}}

\definecolor{aicommentorange}{rgb}{0.85,0.40,0.0}
\newcommand{\aicomment}[1]{\textcolor{aicommentorange}{[AI COMMENT: #1]}}
\definecolor{jiaxingbg}{rgb}{0.78,0.94,0.78}
\definecolor{jiachengbg}{rgb}{0.82,0.88,1.00}

\renewcommand{\aideleted}[1]{}

\renewcommand{\aicomment}[1]{}

\mlsystitlerunning{\sys: Bridging the Gap Between Expressiveness and Efficiency in Attention Masking}

\begin{document}

\twocolumn[
\mlsystitle{\sys: Bridging the Gap Between Expressiveness and Efficiency in Attention Masking}

\mlsyssetsymbol{equal}{*}
\mlsyssetsymbol{intern}{\textdagger}
\mlsyssetsymbol{corresponding}{\textdaggerdbl}
\begin{mlsysauthorlist}
\mlsysauthor{Hongzheng Li}{equal,intern,bupt}
\mlsysauthor{Jiacheng Wan}{equal,jd}
\mlsysauthor{Zhaolong Xing}{jd}
\mlsysauthor{Zhida Jiang}{jd}
\mlsysauthor{Yuanhang Yang}{jd}
\mlsysauthor{Junhua Shu}{jd}
\mlsysauthor{Jiaxing Wang}{jd}
\mlsysauthor{Xupeng Miao}{pku}
\mlsysauthor{Yingxia Shao}{corresponding,bupt}
\mlsysauthor{Zhen Chen}{corresponding,jd}
\mlsysauthor{Ke Zhang}{jd}
\end{mlsysauthorlist}
\mlsysaffiliation{bupt}{Beijing University of Posts and Telecommunications}
\mlsysaffiliation{jd}{JD.com}
\mlsysaffiliation{pku}{Peking University}
\mlsyscorrespondingauthor{\textsuperscript{\textdaggerdbl}Yingxia Shao}{shaoyx@bupt.edu.cn}
\mlsyscorrespondingauthor{Zhen Chen}{chenzhen48@jd.com}
\gdef\compilemaskAuthorNotes{%
  \textsuperscript{*}Equal contribution.
  \textsuperscript{\textdagger}Work done during internship at JD.com.\space%
}

\mlsyskeywords{attention masks, attention kernels, semantic compilation}

\vskip 0.3in

\begin{abstract}

{Modern attention workloads require diverse masks whose shapes depend on per-sequence parameters, yet efficient attention kernels natively support only a small set of mask patterns.
Existing methods for custom masks face an expressiveness-efficiency trade-off.
Predicate-based methods are expressive but evaluate a user-defined predicate on every element at substantial cost, whereas interval-based methods execute efficiently but lack the expressiveness for complex masks.
We present \sys, which bridges this gap by compiling the mask predicate users write into an Execution Descriptor that the attention kernel executes efficiently.
{We design a compiler that automatically lowers a mask predicate into an intermediate representation and generates a mask-specific preprocessing kernel, which builds the descriptor for each batch with a few expressions per row rather than by evaluating the predicate on every element.}
{We further propose Tile-local Records, a compact representation of row visibility within each tile that lets the attention kernel check every element with a few integer operations, so that masking stays efficient even for complex masks.}
Experiments show that \sys~achieves up to $3.3\times$ speedup over FlexAttention and up to $1.2\times$ over FlashMask in single-operator benchmarks, while supporting complex masks that FlashMask cannot express.}
\end{abstract}
]
\thispagestyle{plain}
\printAffiliationsAndNotice{\compilemaskAuthorNotes}  %

\section{Introduction}
\label{sec:introduction}

Attention is the core computation of modern Transformers~\cite{vaswaniAttentionAllYou2017}.
{FlashAttention~\cite{FlashAttention-NueIPS22} has become the de facto attention kernel owing to its efficiency, which comes from computing attention tile by tile without materializing the full score matrix.}
In practice, the set of tokens that each query may attend to varies across tasks, such as generative recommendation~\cite{HSTU_ICML24}, unified multimodal generation~\cite{ICLR2025_12678c39}, video generation~\cite{STA_ICML25}, and preference optimization~\cite{DPO_NeurIPS23}.
Each task encodes this token visibility as an attention mask, which sets the attention scores of invisible query--key pairs to $-\infty$ before the softmax.
The shape of such a mask often depends on per-sequence parameters, such as sequence and segment lengths, and therefore changes from batch to batch.
However, FlashAttention natively supports only a small, fixed set of mask patterns, such as causal and sliding-window masks.
To execute other masks efficiently, practitioners must develop and tune a dedicated attention kernel for each mask pattern, which incurs substantial engineering effort.

FlexAttention~\cite{FlexAttention_MLSys25} lowers this burden with a predicate-based programming model, in which users write a predicate function that specifies whether each element of the attention mask is visible.
The predicate serves both stages of masked execution.
In a preprocessing stage, FlexAttention evaluates it on every query--key pair to build a tile schedule, called \texttt{BlockMask}, which lets the kernel skip tiles whose elements are all masked and bypass the visibility check on tiles whose elements are all visible.
In the attention execution stage, the kernel evaluates the same predicate on each element of the remaining partially masked tiles.
This design is expressive but costly in both stages.
{In preprocessing, building the \texttt{BlockMask} evaluates the predicate on all $N^2$ query--key pairs of a length-$N$ sequence for every batch.
In attention execution, the predicate runs on CUDA cores alongside the softmax, whose throughput on recent GPUs has fallen far behind that of the Tensor Cores used for matrix multiplication~\cite{FlashAttention-3_NeurIPS24,FlashAttention-4_MLSys26}.}

A recent work, FlashMask~\cite{FlashMask_ICLR25}, avoids these costs with an interval-based representation, in which the masked elements of each column are stored as intervals and the kernel checks visibility by comparing element indices against the interval endpoints.
{However, it achieves this efficiency at the cost of expressiveness, because its fixed format holds at most two intervals per column and therefore cannot express masks that need more, which we refer to as \emph{complex masks}.}
{Users must also construct the interval endpoints by hand for each mask.}
Existing approaches therefore face an \emph{expressiveness-efficiency trade-off}, gaining one property only by giving up the other.

We observe that this trade-off is not inherent.
A typical predicate compares the query and key indices with each other and with per-sequence parameters, such as the end of a segment, and combines the outcomes with logical operators.
How it combines these comparisons is fixed once the mask is written, the parameters change only from batch to batch, and only the indices change from element to element, so each part can be handled only as often as it changes instead of all together for every element.
{Handling the parts at different rates means that the representation users write need not be the representation the kernel executes, and translating between the two should be a task for a compiler rather than for the user.}

{To this end, we present \sys, which connects the two representations with a compiler and bridges the gap between expressiveness and efficiency.
Offline, users write the mask as a \texttt{mask\_mod} predicate, the same expressive interface as in FlexAttention, and the compiler analyzes the predicate once and generates a preprocessing kernel specialized for this mask.
Online, for each batch, this kernel takes the per-sequence parameters and produces an Execution Descriptor, from which the attention kernel reads visibility instead of evaluating the predicate.}

{Realizing this pipeline raises two challenges.
The first challenge is how to derive the preprocessing kernel from the predicate, whose Boolean expression hides which keys each row can see.
\sys~addresses it with a symbolic compiler that lowers the predicate into \texttt{MaskSpec}, a Region-based intermediate representation that makes the visible keys of each row explicit, and expresses the flip points of each row, namely the key positions where its visibility changes, as arithmetic expressions of the row index and the per-sequence parameters.
The generated kernel thus evaluates only a few expressions per row for each batch instead of the predicate on every query--key pair.}

{The second challenge is how to design the Execution Descriptor so that the per-element check stays cheap while the descriptor stays compact.
Our key insight is that the attention kernel consumes the mask tile by tile, and within a single tile most rows have only a few flip points even for complex masks.
{\sys~therefore proposes Tile-local Records, which describe each row of a partial tile by its few flip points within that tile, together with Tile-State Bitmaps that classify the tiles.
A Record encodes up to four flip points in a single 32-bit word, and row segments that hold more fall back to a bitmap.
In either case, the kernel checks the visibility of each element with a few integer operations, so masking stays cheap even for complex masks.}}

{With \sys, users write only the predicate, and the compiler generates the preprocessing kernel automatically.
In our production workloads, this reduces the user-written code from approximately 1,300 lines of a hand-written kernel to an 80-line predicate and the development time from days to hours.}

Our contributions are summarized as follows:
\begin{itemize}
    \item {We build a compiler that lowers a FlexAttention-style predicate into \texttt{MaskSpec}, a Region-based intermediate representation, and generates a mask-specific preprocessing kernel.}
    \item {We propose Tile-local Records, which pack the flip points of a row within a tile into one 32-bit word and let the kernel check visibility with a few integer operations regardless of how many intervals the row has.}
    \item {Experiments show that \sys~achieves up to {$3.33\times$ and $3.05\times$} speedups over FlexAttention in the forward and backward passes, respectively, and up to {$1.23\times$} over FlashMask on the masks it can express, while supporting complex masks beyond its reach.}
\end{itemize}

\section{{Background and Motivation}}
\label{sec:background}

\begin{figure}[!t]
\centering
\includegraphics[width=\columnwidth]{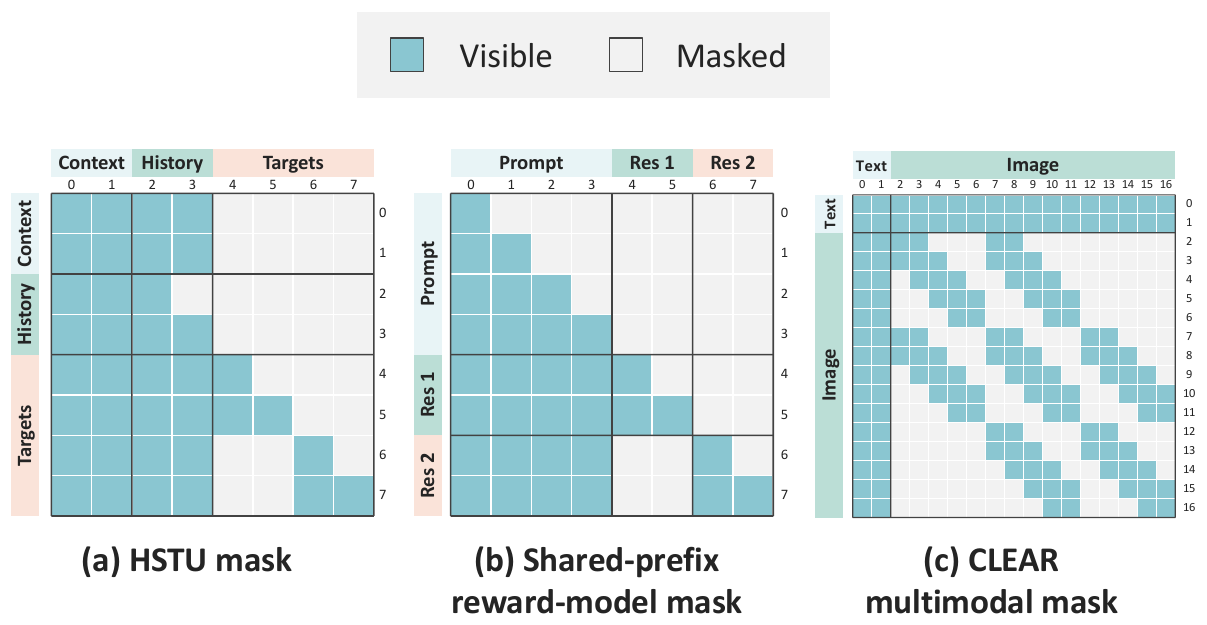}
\vspace{-15pt}
\caption{{Examples of attention masks for three workloads.}}
\label{fig:region_interval}
\end{figure}

\subsection{Preliminary}
\label{subsec:bg_execution}

\paragraph{{Attention masking and tiling.}}
Given query, key, and value matrices $Q,K,V\in\mathbb{R}^{N\times d}$, where $N$ is the sequence length and $d$ is the head dimension, attention forms the score matrix $S$ and output $O$ as
\begin{equation}
\label{eq:masked_attention}
\begin{aligned}
S &= QK^{\top}/\sqrt{d}, \qquad O=\operatorname{softmax}(S+M)V,\\
M_{i,j} &=
\begin{cases}
0 & \text{if key } j \text{ is visible to query } i,\\
-\infty & \text{otherwise}.
\end{cases}
\end{aligned}
\end{equation}
{The mask determines which keys each query attends to and may depend on per-sequence parameters.}
{Figure~\ref{fig:region_interval} shows three such masks, namely the HSTU mask for generative recommendation~\cite{HSTU_ICML24}, the shared-prefix reward-model mask for preference optimization~\cite{DPO_NeurIPS23,DPO-PrefixCache_NeurIPS24-WS}, and the CLEAR mask for unified multimodal generation~\cite{CLEAR_NeurIPS25}.}

FlashAttention~\cite{FlashAttention-NueIPS22} computes attention tile by tile without materializing the full score matrix.
Fully masked tiles can be skipped, fully visible tiles require no masking, and partially masked tiles require element-wise visibility checks.
{Supporting a custom mask therefore involves classifying tiles and resolving visibility within partial tiles.}

\begin{figure*}[t]
\centering
\includegraphics[width=0.9\textwidth]{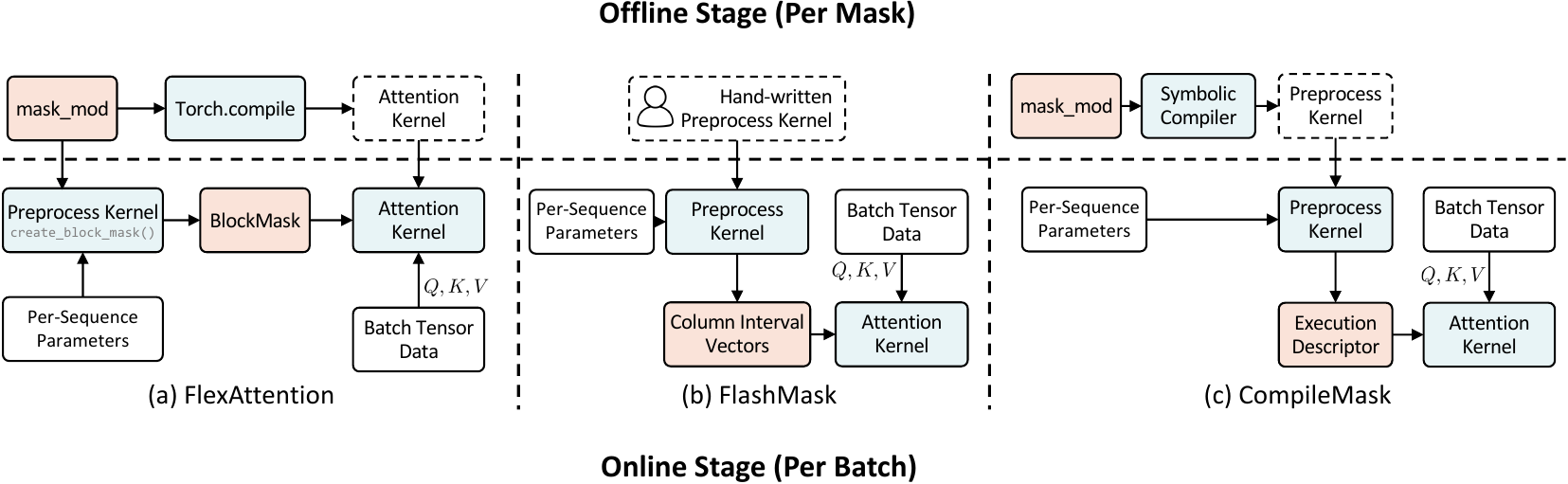}
\caption{{Pipelines of (a) the predicate-based method, represented by FlexAttention, (b) the interval-based method, represented by FlashMask, and (c) \sys.
The upper row shows the offline stage, performed once per mask, and the lower row shows the online stage, performed once per batch.
Pink blocks denote mask representations, green blocks denote computational components.}}
\label{fig:overview}
\end{figure*}

\paragraph{Predicate-based masking.}
FlexAttention~\cite{FlexAttention_MLSys25} accepts {a \emph{mask predicate}, an element-wise Boolean function \texttt{mask\_mod}$(i,j)$ that determines whether key $j$ is visible to query $i$ and may read per-sequence parameters}, providing an expressively complete interface.
{Its flexibility comes with overhead in both preprocessing and attention execution, because the predicate flows unchanged into both stages, as Figure~\ref{fig:overview}(a) shows.}
During preprocessing, \texttt{create\_block\_mask} evaluates the predicate across query--key pairs to identify tiles that can be skipped or processed without masking.
{This construction can be costly and must be repeated for every batch due to the changing of the per-sequence parameters.}
During attention execution, the kernel still evaluates the user-defined predicate for each element within partially masked tiles.
{The predicate therefore adds arithmetic and logical operations to the attention kernel, making online visibility checks expensive even after the \texttt{BlockMask} has been constructed.}

\paragraph{Interval-based masking.}
FlashMask~\cite{FlashMask_ICLR25} represents a mask using four per-column boundary vectors {(\texttt{LTS}, \texttt{LTE}, \texttt{UTS}, \texttt{UTE})}{, as Figure~\ref{fig:overview}(b) shows}, where \texttt{LTS} and \texttt{LTE} specify the start and end of a masked interval in the lower triangular part of each column and \texttt{UTS} and \texttt{UTE} specify those in the upper triangular part.
During preprocessing, it computes \texttt{MinMax} summaries by taking the minimum and maximum of each of these four vectors within each column block.
The \texttt{MinMax} summaries are compared against a query tile's row range to classify the tile as fully masked, fully visible, or partially masked.
Within partially masked tiles, the attention kernel checks the column intervals to determine element-wise visibility.
This representation supports efficient execution, but its fixed number of boundary vectors limits expressiveness.
{Each pair of vectors describes only one contiguous masked interval, so the format cannot represent complex masks, whose columns contain more than two masked intervals, such as column 9 of the CLEAR~\cite{CLEAR_NeurIPS25} mask in Figure~\ref{fig:region_interval}(c).
Moreover, users must derive these vectors by hand for each mask.}
{Predicate-based masking thus retains expressiveness at the cost of efficiency, whereas interval-based masking gains efficiency at the cost of expressiveness, which is the expressiveness-efficiency trade-off that \sys~sets out to avoid.}

\subsection{{Motivation and Challenges}}
\label{subsec:bg_motivation}
{A mask predicate decides visibility by comparing the query and key indices with each other and with per-sequence parameters, and by combining the outcomes with logical operators.
In the HSTU mask of Figure~\ref{fig:region_interval}(a), for example, the context and history terms state that query $i$ may attend to key $j$ if $i < c$ and $j < b$, or if $c \le i < b$ and $j \le i$, where $c$ and $b$ are the ends of the context and history segments.
Such a predicate has three ingredients that change at very different rates.
Its \emph{Boolean structure}, namely which comparisons it makes and how it combines them, is fixed once the mask is written.
Its per-sequence parameters, here $c$ and $b$, change from batch to batch, and only the indices $i$ and $j$ change from element to element.
This motivates handling each ingredient only as often as it changes, with the Boolean structure analyzed once per mask, the per-sequence parameters applied once per batch, and only the index comparisons left for each element.}

{The two challenges that this decoupling raises are already visible in the HSTU example.
Deriving what must be computed for each batch is hard to automate, because the predicate mixes comparisons on $i$ with comparisons on $j$ in one Boolean expression, so where the visibility of a row changes along the keys is only implicit in it.}
{Representing the per-batch result is equally delicate, because the kernel must check each element cheaply while the representation stays compact.}
Storing all visible intervals of each row (column), which we call \emph{per-row intervals}, lifts the interval limit of FlashMask, but the check of each element then grows with the number of intervals, which reaches tens to hundreds in long-sequence recommendation~\cite{BlossomRec_WWW26,KSA_arXiv26} and video generation~\cite{STA_ICML25,GNA_CVPR26}.
{Section~\ref{subsec:method_overview} builds the pipeline of \sys~on the above motivation, and Sections~\ref{subsec:mask_compilation} and~\ref{subsec:descriptor_execution} address the two challenges, respectively.}

\section{Methodology}
\label{sec:methodology}

\subsection{{The \sys~Pipeline}}
\label{subsec:method_overview}

{Following the motivation in Section~\ref{subsec:bg_motivation}, \sys~handles each ingredient of a mask predicate at the level at which it changes.}
{It analyzes the Boolean structure once per mask and reduces it, for each query row, to a few key positions where visibility changes, {which we call \emph{flip points}}.}
{It computes these positions from the per-sequence parameters once per batch, so that each element of a partial tile in every attention layer only needs a lightweight check of its key index against them.}

{Figure~\ref{fig:overview}(c) shows the pipeline of \sys.}
Offline, users write the mask as a \texttt{mask\_mod} predicate, the same interface as in FlexAttention~\cite{FlexAttention_MLSys25}.
{Our symbolic compiler, detailed in Section~\ref{subsec:mask_compilation}, analyzes this predicate, rewrites its visibility rules as arithmetic expressions of the row index and the per-sequence parameters, and generates a preprocess kernel specialized for the mask.
Online, for each batch, the preprocess kernel takes the per-sequence parameters of the batch, evaluates these expressions, and outputs an Execution Descriptor.
This step thus evaluates a few expressions per row instead of the predicate on every query--key pair.
The Execution Descriptor is carefully designed to keep attention execution lightweight, as Section~\ref{subsubsec:semantic_descriptor} details.}

{Figure~\ref{fig:overview} contrasts this pipeline with the two existing approaches.
In the predicate-based method of Figure~\ref{fig:overview}(a), the predicate flows unchanged into both online components, so the mask is evaluated again for every batch and in every layer.
The interval-based method of Figure~\ref{fig:overview}(b) keeps its online stage as light as that of \sys, but users must hand-write its offline preprocess kernel, whose output is limited to two intervals per column.
\sys~replaces this hand-written kernel with a compiler that accepts the same predicate as the predicate-based method and emits an Execution Descriptor with no fixed interval limit, which decouples the representation users write from the one the kernel executes and thus combines the expressiveness of the former method with the efficiency of the latter.}
{Section~\ref{subsec:mask_compilation} describes how the compiler lowers a predicate into {an intermediate representation} and generates the preprocess kernel, and Section~\ref{subsec:descriptor_execution} describes the Execution Descriptor and how the attention kernel uses it.}

\subsection{{Offline Mask Compilation}}
\label{subsec:mask_compilation}

{Users write the mask as a \texttt{mask\_mod} predicate, the interface of FlexAttention~\cite{FlexAttention_MLSys25}, which takes the batch, head, query, and key indices and returns whether the key is visible to the query.}
\begin{lstlisting}[style=signature]
def mask_mod(batch_idx: int, head_idx: int, q_idx: int, kv_idx: int) -> bool
\end{lstlisting}
{Section~\ref{subsec:method_overview} requires the compiler to reduce such a predicate, for each query row, to its flip points, expressed as arithmetic expressions of the row index and the per-sequence parameters.
A predicate does not expose these positions directly, because it mixes comparisons on the query index and comparisons on the key index within one Boolean expression.
\sys~therefore analyzes the source code of the predicate symbolically and lowers it into a Region-based Intermediate Representation (IR), which we call \texttt{MaskSpec}.}

\begin{figure}[!t]
    \centering
    \includegraphics[width=\linewidth]{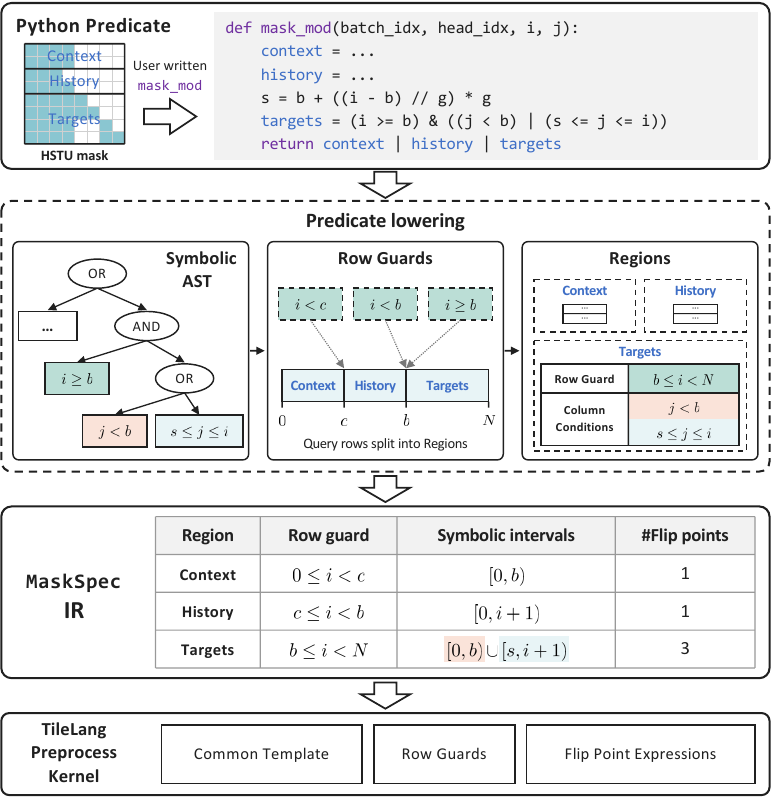}
    \vspace{-15pt}
    \caption{{Compilation of the HSTU mask from a Python predicate into a TileLang preprocess kernel.
    $c$, $b$, $N$, and $g$ are per-sequence parameters.}}
    \label{fig:symbolic_compilation}
    \end{figure}

\paragraph{{Lowering predicates to \texttt{MaskSpec}.}}
\label{subsubsec:mask_specification}
\label{subsubsec:symbolic_lowering}
{The compiler lowers a predicate by turning a Boolean test on each query--key pair into intervals of keys for each row, as illustrated for the HSTU mask in Figure~\ref{fig:symbolic_compilation}.
It parses the predicate with Python's \texttt{ast.parse} and inlines its intermediate variables.
Comparisons that involve only the query index become row guards, and each comparison on the key index is solved for the key index into an interval, such as $j<b$ into $[0,b)$.
An OR between comparisons on the key index becomes a union of their intervals, and a loop becomes the union of a family of intervals whose count is a per-sequence parameter.
}

{The result is a \texttt{MaskSpec}, which represents a mask as a set of Regions $R_1,\dots,R_G$.
Each Region $R_g$ pairs a row guard $\operatorname{when}_g(i)$ with a rule built from symbolic intervals $[L(i),U(i))$, whose endpoints are arithmetic expressions of the row index $i$ and the per-sequence parameters.
The visible keys $\mathcal{V}(i)$ of row $i$ are given by the rule of the Region $\rho(i)$ whose guard holds,
\begin{equation}
\label{eq:region_semantics}
\mathcal{V}(i) = \bigcup_{k=1}^{n_{\rho(i)}} \left[L_{\rho(i),k}(i),\,U_{\rho(i),k}(i)\right),
\end{equation}
where $n_{\rho(i)}$ is the number of intervals in the rule of $\rho(i)$.
{Once overlapping or adjacent intervals are merged, $\mathcal{V}(i)$ becomes a union of disjoint, non-adjacent intervals, and their endpoints give the flip points of the row.}
Because the endpoints stay symbolic, the compilation runs once per mask rather than once per batch.}

\paragraph{{Preprocess kernel generation.}}
\label{subsubsec:preprocess_generation}
{The compiler generates the preprocess kernel by filling a fixed TileLang template with the mask-specific parts, namely the row guards and the flip point expressions.
At runtime, the kernel evaluates the flip point expressions with the per-sequence parameters of each batch and builds the Execution Descriptors of both directions, as Section~\ref{subsubsec:semantic_descriptor} describes.
{For a row segment with more than four flip points, which a Record cannot hold, the kernel writes the exact visibility bitmap of the segment instead, as Section~\ref{subsubsec:element_check} describes.}
For the backward direction, it transposes the forward visibility of each tile instead of lowering the predicate again, so users write one predicate for both passes.}

\begin{figure*}[!t]
\centering
\includegraphics[width=0.9\textwidth]{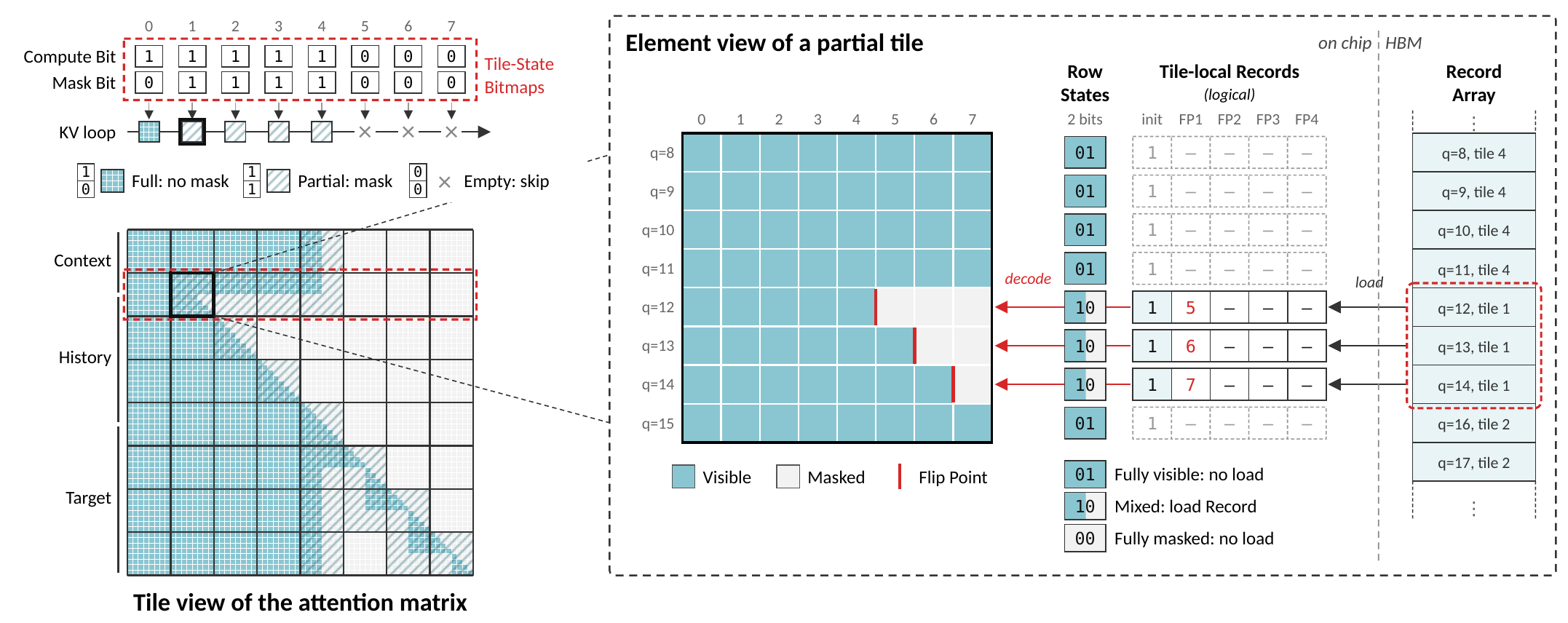}
\caption{{Descriptor-guided attention execution.
The Tile-State Bitmaps classify each tile, and within a partial tile the Row States and the logical Tile-local Records determine the visibility of each element.}}
\label{fig:descriptor_guided_execution}
\end{figure*}

\subsection{{Online Execution with Tile-Local Records}}
\label{subsec:descriptor_execution}

{Interval-based masking, introduced in Section~\ref{sec:background}, executes masks efficiently online, but the existing method~\cite{FlashMask_ICLR25} holds at most two intervals in its fixed format and cannot express complex masks, which require more than two intervals.
A natural extension would be to store all intervals of each query row, up to $K$ per row, and we call this naive method \emph{per-row intervals}.
Its attention kernel checks every element of a partial tile against all intervals of the row, so both the per-element compute and the registers grow with $K$.}

{\sys avoids this cost with Tile-local Records, which describe the visibility of a row within one tile by its flip points, namely the columns at which the visibility of the row changes.
We observe that, even for complex masks, most rows contain only a few flip points within a tile, so the per-element cost of masking no longer depends on how many intervals the row has.
For masks that are not complex, its cost stays of the same order as with per-row intervals.
The Records form one part of the Execution Descriptor, which also classifies tiles and rows so that the attention kernel reads Records only where they are needed.
We first describe how the attention kernel uses the Execution Descriptor at the tile level and then how it checks the visibility of each element with the Records.}

\subsubsection{{Tile-Level Execution}}
\label{subsubsec:semantic_descriptor}

{{At runtime, the preprocess kernel generated in Section~\ref{subsec:mask_compilation} evaluates the flip point expressions with the per-sequence parameters of each batch and constructs the Execution Descriptors for the forward and backward passes.}
Each Execution Descriptor is composed of \textit{Tile-State Bitmaps}, which classify each tile, together with \textit{Row States} and {the \textit{Tile-local Records} of the partial tiles}.}
{Figure~\ref{fig:descriptor_guided_execution} shows these Execution Descriptors and how the attention kernel uses them.}

To skip empty tiles and avoid masking in full tiles, the attention kernel needs tile-level visibility information.
We encode this information in the Tile-State Bitmaps, using only two bits per tile.
{The compute bit indicates whether any element is visible, while the mask bit indicates whether the tile contains both visible and masked elements.}
Together, they distinguish empty, full, and partial tiles, as shown in Figure~\ref{fig:descriptor_guided_execution}.
{The attention kernel uses these bits to select the execution path of each tile.
Empty tiles are skipped without loading their key and value tiles or performing matrix multiplication.
Full tiles follow the standard attention computation without masking, while partial tiles require element-wise visibility checks using the {Tile-local Records}.}

{{Even within a partial tile, individual rows can be entirely visible or entirely masked.}
Loading a visibility description for every row would therefore spend memory traffic on rows that need none.
Therefore, we use a two-bit Row State for each row in the partial tile, which marks the row as fully visible (01), fully masked (00), or mixed (10), as illustrated in Figure~\ref{fig:records_encoding}{, and sixteen Row States share one 32-bit word}.
This ensures that a partial tile loads Records only for its mixed rows.
{For fully visible and fully masked rows, the attention kernel uses a predefined constant instead of loading a Record, and the masking step treats all rows uniformly to perform the visibility check, which will be described in Section~\ref{subsubsec:element_check}.}}
{The Row States and the Records reside in global memory, each stored in the order in which the attention kernel visits the partial tiles, so the kernel reads them sequentially without any index.
{{The Records form a \emph{Record Array}, in which each mixed row segment (i.e., the part of a row within one tile), takes one 32-bit Record, whose encoding Section~\ref{subsubsec:element_check} describes.}
Because a row segment is mixed only when it contains a flip point of the row, the Records of a row take no more words than its per-row intervals, so the descriptor stays small, as Table~\ref{tab:records_cost} shows.}}
{
The descriptor is built once per batch and reused across all attention layers that share the mask.}

{
\paragraph{Row States and Records prefetching.}
{Loading the Row States and Records of a partial tile only when the tile is masked would place a global memory access on the critical path of every partial tile.
The attention kernel therefore loads them for the next partial tile right after it masks the current one, as the kernel loop in Appendix~\ref{subsec:app_kernel_loop} shows, so the loads overlap with the softmax of the current tile and the $QK^\top$ of the next tile.}}

\subsubsection{{Element-Level Visibility Check}}
\label{subsubsec:element_check}

\begin{figure}[!t]
\centering
\includegraphics[width=0.85\columnwidth]{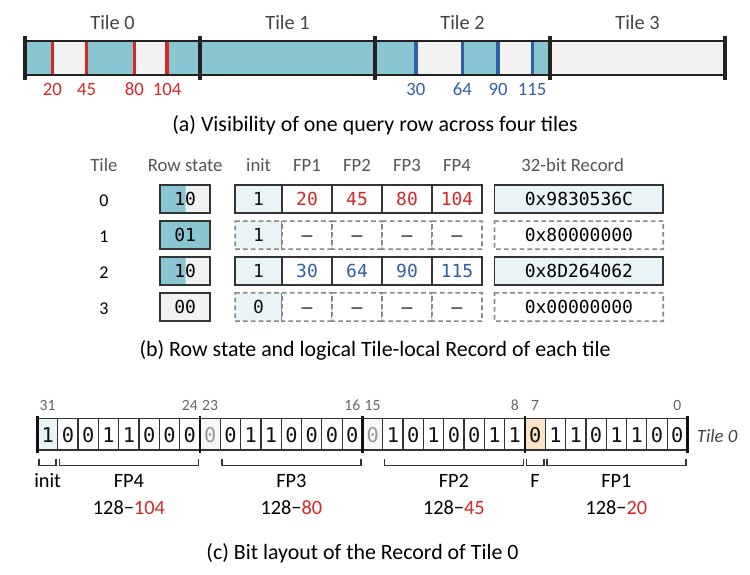}
\vspace{-5pt}
\caption{{Tile-local Records of one query row across four tiles of 128 columns.
F marks the bitmap fallback.}}
\label{fig:records_encoding}
\end{figure}

{A Record is a single 32-bit word, as shown in Figure~\ref{fig:records_encoding}.
It holds $\mathrm{init}$, the visibility of the row at the first column of the tile, and up to four flip points within the tile.}
{Within a partial tile, the visibility of column $c$ of a mixed row $i$ with flip points $p_{i,1},\dots,p_{i,m}$, where $m\le 4$, is the initial visibility flipped once for every flip point at or before $c$,}
\begin{equation}
\label{eq:flip_parity}
{\operatorname{visible}(i,c) = \mathrm{init}_i \oplus \bigoplus_{k=1}^{m} \left[\, c \ge p_{i,k} \,\right].}
\end{equation}
{Because the attention kernel checks the visibility of every element of a partial tile, this check must be lightweight and take only a few instructions.}
{Evaluating Equation~\ref{eq:flip_parity} directly would unpack the four flip points into separate registers and spend one comparison and one exclusive or per flip point on every element.
It would thus add both instructions per element and registers per row to a kernel whose registers are already occupied by the accumulators.}
{We therefore encode the flip points so that the kernel checks the visibility of an element with a few integer operations on the packed Record.
Each flip point at column $p$ is stored as $128-p$ in its own byte, as shown in Figure~\ref{fig:records_encoding}.
With this offset, the sum $128-p+c$ reaches 128 exactly when $c \ge p$, so the comparison of $c$ with $p$ becomes the top bit of the byte after adding $c$.
Building on this encoding, we check the visibility of column $c$ from a Record $w$ with}
\begin{equation}
\label{eq:decode}
{\textup{\textsc{Decode}}(w,c) = \big(\big((w + c\,u) \mathbin{\&} 128u\big)\cdot u\big) \gg 31,}
\end{equation}
{where $u=\mathtt{0x01010101}$ copies $c$ into all four bytes, so that one addition compares $c$ with all four flip points at once.
This check takes only five integer operations regardless of how many intervals the row has, and Theorem~\ref{thm:record_decoding} shows that it is exact.}

\begin{theorem}[Record decoding]
\label{thm:record_decoding}
{Let $w$ be the Record of a row segment with at most four flip points, and let $0\le c<128$ be a column of the tile.
With 32-bit unsigned arithmetic, $\textup{\textsc{Decode}}(w,c)$ in Equation~\ref{eq:decode} equals $\operatorname{visible}(i,c)$ in Equation~\ref{eq:flip_parity}.
For the constant Records $\mathtt{0x80000000}$ and $0$, it yields 1 and 0 for every column.}
\end{theorem}

{Appendix~\ref{sec:app_records_theory} gives the proof.
The kernel then replaces the scores of invisible elements with $-\infty$ before applying the online softmax.}

{A 32-bit Record has room for at most four flip points.}
{{When a row segment has more than four flip points, its Record instead refers to a dense bitmap with one bit per column of the segment, which stores the exact visibility of each element of the segment.}
Bit 7 of the Record distinguishes the two cases.
If this bit is zero, the Record holds flip points, and otherwise its remaining bits hold the offset of the bitmap in the Record Array.
When the kernel applies the mask, it reads the bitmap at this offset instead of decoding the Record.
The fallback applies only to the row segments with more than four flip points, so it keeps the mask exact without changing how any other row is encoded.}

\begin{table}[t]
\centering
\caption{{Cost of masking in the attention kernel, where $K$ is the largest number of intervals of a row and $P_i$ is the number of partial tiles of row $i$.}}
\label{tab:records_cost}
\small
\setlength{\tabcolsep}{4pt}
\resizebox{\linewidth}{!}{%
\begin{tabular}{cccc}
\toprule
 & & {Per-row intervals} & {Tile-local Records} \\
\midrule
\multirow{2}{*}{\begin{tabular}[c]{@{}c@{}}Masking\\compute\end{tabular}} & {Operations per element} & {$O(K)$} & {$O(1)$} \\
 & {Registers per row} & {$O(K)$} & {$O(1)$} \\
\midrule
\multirow{2}{*}{\begin{tabular}[c]{@{}c@{}}Descriptor\\loads\end{tabular}} & {Intervals or Records} & {$O(K)$} & {$O(K)$} \\
 & {Row States} & {--} & {$O(P_i/16)$} \\
\bottomrule
\end{tabular}%
}
\end{table}

{Table~\ref{tab:records_cost} summarizes the resulting costs, and Appendix~\ref{sec:app_records_theory} gives their derivations.
With Tile-local Records, the operations per element and the registers per row of masking stay constant instead of growing with $K$, so complex masks no longer slow down the check of each element.
{At the same time, the descriptor data stay of the same order as with per-row intervals apart from a small amount of Row States, whose loads are hidden behind computation by prefetching.}}

\section{Experiments}
\label{sec:experiment}

\begin{figure*}[!t]
\centering
\includegraphics[width=\textwidth]{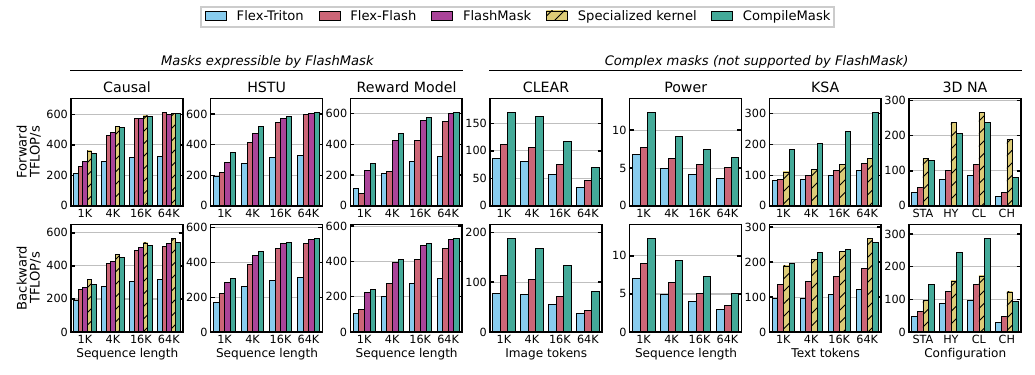}
\vspace{-10pt}
\caption{{Effective forward and backward throughput of a single attention operator, excluding mask preprocessing.
The left three masks can be expressed by FlashMask, and the right four are complex masks that it cannot express.
The specialized kernel is Native FA for Causal, the official KSA kernel for KSA, and NATTEN for 3D NA.
Higher is better.}}
\label{fig:operator_fwd_bwd}
\end{figure*}

\subsection{{Experimental Setup}}
\label{subsec:exp_setup}

\paragraph{{Environment.}}
{All experiments run on NVIDIA Hopper GPUs with BF16 precision, PyTorch 2.13.0, and CUDA 13.0.
\sys~uses TileLang 0.1.9 for preprocessing kernels and CUTLASS DSL 4.8.0.dev0 for attention kernels.
{Appendix~\ref{sec:app_settings} summarizes the workload parameters of every experiment.}}

\paragraph{{Workloads.}}
{We evaluate seven attention masks in two groups.
{The first group contains masks that FlashMask can express, namely \textbf{Causal}~\cite{vaswaniAttentionAllYou2017}, \textbf{HSTU}~\cite{HSTU_ICML24} from generative recommendation, and the shared-prefix reward-model mask~\cite{DPO-PrefixCache_NeurIPS24-WS}, denoted as \textbf{RM}.}
{The second group contains complex masks, whose columns contain more than two masked intervals, so FlashMask cannot express them.}
It includes \textbf{CLEAR}~\cite{CLEAR_NeurIPS25} for text--image attention, the \textbf{Power} pattern of BlossomRec~\cite{BlossomRec_WWW26}, Kwai Summary Attention~\cite{KSA_arXiv26}, denoted as \textbf{KSA}, and three-dimensional neighborhood attention over video tokens, denoted as \textbf{3D NA}.
{For 3D NA, we use one configuration from STA~\cite{STA_ICML25} and the Hunyuan, Cosmos-low, and Cosmos-high configurations from GNA~\cite{GNA_CVPR26}, denoted as STA, HY, CL, and CH, respectively.}
{Figure~\ref{fig:app_mask_shapes} in the appendix illustrates the shape of each mask.}
Lengths count image tokens for CLEAR, text tokens for KSA, and all tokens for the other masks, where K denotes 1,024.
{All methods receive identical inputs and mask metadata.}}

\paragraph{{Baselines.}}
{We compare against \textbf{FlexAttention}~\cite{FlexAttention_MLSys25} and \textbf{FlashMask}~\cite{FlashMask_ICLR25}.
FlexAttention uses either its Triton backend or its FLASH backend, denoted as \textbf{Flex-Triton} and \textbf{Flex-Flash}, respectively, and both backends use a BlockMask granularity of $128\times128$.
On Hopper GPUs, the FLASH backend dispatches to {the SM90 CuTeDSL kernels of FlashAttention-4~\cite{FlashAttention-4_MLSys26}}, on which the attention kernels of \sys~are also built.
In single-operator benchmarks, FlashMask uses its official CuTeDSL implementation for SM90\footnote{\url{https://github.com/PaddlePaddle/flash-attention/tree/b492328/flashmask/flash_mask/cute}} with PaddlePaddle 3.3.0.
Since this implementation is not integrated into PyTorch, end-to-end training uses a PyTorch port of the same kernels.
For three masks, we also compare against kernels written specifically for them, which serve as references for the practical performance ceiling.
These specialized kernels are {FlashAttention-4 with built-in causal masking} for Causal, denoted as \textbf{Native FA}, the \textbf{official kernel of KSA}~\cite{KSA_arXiv26}{\footnote{\url{https://github.com/Kuaishou-OneRec/KSA/tree/9998565}}}, and the Hopper kernel of \textbf{NATTEN}~\cite{GNA_CVPR26}{\footnote{\url{https://github.com/SHI-Labs/NATTEN}, version 0.21.7.}} for 3D NA.
}

\paragraph{{Metrics.}}
{For a single operator, we report effective throughput, which divides the attention FLOPs over visible query--key pairs by the measured time.
Following FlashAttention~\cite{FlashAttention-2_ICLR24}, we count the backward pass as 2.5 times the forward-pass FLOPs to include score recomputation.
For end-to-end training, we report throughput in thousands of tokens per second, denoted as kTok/s.
{Every reported time is averaged over repeated measurements after warmup.}
We validate attention outputs and gradients with respect to queries, keys, and values against dense masked attention, as reported in Appendix~\ref{sec:app_correctness}.}

\begin{table}[!t]
\centering
\caption{{Mask preprocessing time in milliseconds and the speedup of \sys~over FlexAttention.
Flex denotes FlexAttention, and CM denotes \sys.}}
\label{tab:preprocess_hopper}
\footnotesize
\setlength{\tabcolsep}{4pt}
\setlength{\tabcolsep}{3pt}
\resizebox{\columnwidth}{!}{%
\begin{tabular}{ccccccccccc}
\toprule
& \multicolumn{2}{c}{HSTU} & \multicolumn{2}{c}{CLEAR} & \multicolumn{2}{c}{Power} & \multicolumn{2}{c}{KSA} & \multicolumn{2}{c}{3D NA} \\
\cmidrule(lr){2-3} \cmidrule(lr){4-5} \cmidrule(lr){6-7} \cmidrule(lr){8-9} \cmidrule(l){10-11}
Method & 4K & 16K & 4K & 16K & 4K & 16K & 4K & 16K & STA & CL \\
\midrule
Flex & 2.01 & 7.24 & 6.64 & 24.64 & \textbf{0.46} & 4.34 & 1.64 & 11.83 & 45.98 & 23.17 \\
CM & \textbf{0.09} & \textbf{0.25} & \textbf{0.83} & \textbf{2.47} & 0.57 & \textbf{0.79} & \textbf{1.21} & \textbf{4.62} & \textbf{9.87} & \textbf{6.61} \\
\midrule
Speedup & 23.3$\times$ & 28.5$\times$ & 8.1$\times$ & 10.0$\times$ & 0.8$\times$ & 5.5$\times$ & 1.4$\times$ & 2.6$\times$ & 4.7$\times$ & 3.5$\times$ \\
\bottomrule

\end{tabular}}
\vspace{-15pt}
\end{table}

\subsection{{Single-Operator Performance}}
\label{subsec:kernel_performance}
\label{subsubsec:operator_protocol}
\label{subsubsec:operator_integrated}

{We first measure the forward and backward passes of a single attention operator.
All benchmarks use a head dimension of 128 and fixed-length sequences.
For the masks evaluated from 1K to 64K, each batch contains 128K tokens in eight attention heads, so the batch size decreases from 128 to 2 as the length grows.
The 3D NA configurations use 24 heads and one or two videos per batch.}

\paragraph{{Masks expressible by FlashMask.}}
{Figure~\ref{fig:operator_fwd_bwd} shows that \sys~achieves the highest forward and backward throughput of all methods on HSTU and RM at all four lengths.
Its advantage is largest at short sequences, where at 1K it reaches up to {$3.33\times$} the forward throughput of Flex-Flash and {$1.23\times$} that of FlashMask.
At 64K, \sys~and FlashMask converge to within 2\% of each other in both passes.}

{\sys~gains most at short sequences because its visibility check costs less per element.
Only partially visible tiles need the check, and these tiles make up a larger share of the work at short sequences.
FlexAttention evaluates the user predicate on every element of these tiles, whereas \sys~decodes a Tile-local Record with a fixed number of instructions per element.
\sys~also matches or exceeds FlashMask, whose four-vector representation is designed specifically for these masks, through more efficient engineering in its kernel implementation.}

\paragraph{{Complex masks.}}
{FlashMask cannot express any of the four complex masks, because its fixed four-vector representation holds at most two masked intervals per column.
CLEAR already needs three or more masked intervals in some columns, and Power, KSA, and 3D NA {need from tens to more than a thousand visible intervals per row or column}.
{On these masks, \sys~is {$1.26$--$2.45\times$ and $1.36$--$2.28\times$} faster than Flex-Flash in the forward and backward passes, respectively.}
Because complex masks leave many tiles partially visible, the constant per-element cost of Tile-local Records matters more than on the first group, which explains the larger gains over FlexAttention.}

\paragraph{{Comparison with specialized kernels.}}
{{The specialized kernels serve as performance references for a general mask compiler.}
{On the causal mask, \sys~stays within 5\% of the forward throughput and within 9\% of the backward throughput of Native FA, which closely approaches this reference.}
{{\sys~is up to $1.97\times$ faster than the official kernel in the forward pass of KSA and $1.51$--$1.67\times$ faster than NATTEN in the backward pass of 3D NA on STA, HY, and CL.
This is because both specialized kernels evaluate the original complex predicate for every element during attention execution, which lowers the active ratio of the Tensor Cores.
\sys~instead resolves the predicate logic before attention execution and only decodes a Record with a few bit operations per element, so the visibility check stays lightweight.}}}
On CH, \sys~is slower than NATTEN, which uses tiling specialized for 3D neighborhood attention, forming each tile from a contiguous region of the video grid rather than a contiguous segment of the flattened sequence, so visible pairs fill 66\% of its non-empty tiles, compared with 19\% for \sys.
However, this efficiency relies on neighborhood-specific layout and kernel design, whereas \sys~supports different masks without requiring such specialized implementations.
These results show that \sys~stays close to specialized kernels and outperforms them in several configurations, without requiring a hand-written kernel for each mask.

\paragraph{Preprocess Overhead.}
{Mask preprocessing runs once per training step, and Table~\ref{tab:preprocess_hopper} reports its GPU time.
For \sys, preprocessing constructs the Tile-State Bitmaps and the Tile-local Records for both passes from per-sequence parameters, whereas FlexAttention constructs the BlockMask by evaluating the predicate on every element.
{The Triton and FLASH backends of FlexAttention share the same preprocessing, so Table~\ref{tab:preprocess_hopper} reports a single FlexAttention entry.}
\sys~takes {$1.4$--$28.5\times$} less preprocessing time than FlexAttention in all configurations except Power at 4K, where both take less than 0.6~ms.
Because one preprocessing result serves every attention layer, we also compare the share of preprocessing in the attention time of a training step, $r_{\mathrm{pre}} = T_{\mathrm{pre}}/(T_{\mathrm{pre}} + L\,T_{\mathrm{attn}})$, where $T_{\mathrm{pre}}$ is the preprocessing time, $T_{\mathrm{attn}}$ is the forward and backward attention time of one layer, and $L$ is the number of attention layers.
We set $L=36$ as a representative value, following the configuration of Qwen3-8B~\cite{Qwen3_arxiv25}.
{On CLEAR at 16K, preprocessing accounts for 4.5\% and 5.8\% of the attention time for Flex-Triton and Flex-Flash, respectively, but only 1.1\% for \sys.}}

\begin{figure}[!t]
\centering
\includegraphics[width=\columnwidth]{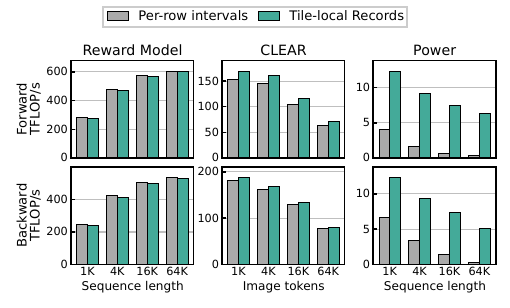}
\vspace{-20pt}
\caption{{Throughput of per-row intervals and Tile-local Records on masks whose rows hold increasingly many intervals.}}
\label{fig:real_masks_vs_interval}
\end{figure}

\begin{figure}[!t]
\centering
\includegraphics[width=0.85\columnwidth]{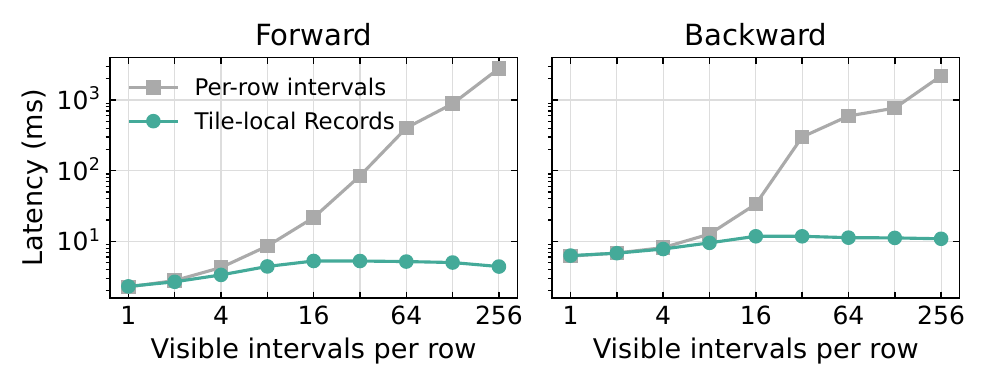}
\caption{{Latency of per-row intervals and Tile-local Records under different numbers of visible intervals per row.}}
\label{fig:checkerboard}
\end{figure}

\subsection{{Tile-local Records versus Per-Row Intervals}}
\label{subsec:records_vs_intervals}

{To isolate the effect of the visibility encoding, we compare \sys~with the per-row interval design described in Section~\ref{subsec:descriptor_execution}, which keeps the same attention kernel and tile skipping but checks every element against all visible intervals of its row.}

{Figure~\ref{fig:real_masks_vs_interval} compares the two encodings on masks whose rows hold increasingly many intervals.
On RM, whose rows have few intervals, the two encodings perform within {3\%} of each other.
As rows hold more intervals, Tile-local Records pull ahead of per-row intervals, by up to {$1.12\times$} on CLEAR and by up to {$14.6\times$ and $18.3\times$} in the forward and backward passes on Power.}

{
To further study how the number of intervals per row affects the two encodings, we conduct experiments on an additional mask, named the checkerboard mask, {in which visible and masked square cells of equal size alternate along both axes, so that each query row alternates between visible and masked intervals of equal length, as visualized in Figure~\ref{fig:app_checkerboard_shapes} in the appendix.}
We adjust the number of visible intervals per row from {1} to 256 on sequences of about 4K tokens.
As Figure~\ref{fig:checkerboard} shows, the latency of per-row intervals grows with the number of intervals, whereas the forward latency of Tile-local Records stays between {2.3 and 5.3}~ms 
because each tile-local representation encodes visibility only within one tile. 
With 256 intervals per row, Tile-local Records are {$635\times$ and $202\times$} faster than per-row intervals in the forward and backward passes.
{Tile-local Records therefore cost little when rows have few intervals and avoid the per-element cost that per-row intervals pay on complex masks.}}

\begin{figure}[!t]
\centering
\includegraphics[width=\columnwidth]{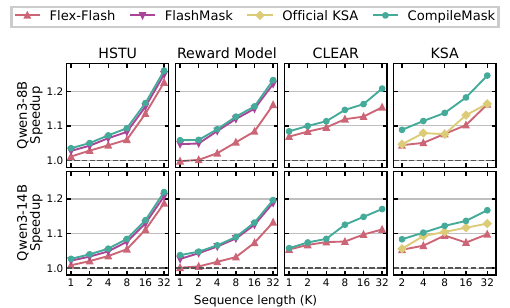}
\vspace{-15pt}
\caption{{End-to-end training throughput of Qwen3 models, normalized to Flex-Triton.}}
\vspace{-5pt}
\label{fig:qwen3_hstu_end_to_end}
\end{figure}

\subsection{{End-to-End Performance}}
\label{subsec:end_to_end}

{{We conduct end-to-end model training on Qwen3-8B and Qwen3-14B~\cite{Qwen3_arxiv25}} with Hugging Face Transformers on a single GPU with BF16, AdamW, \texttt{torch.compile}, and activation checkpointing.
Here we set batch size to one without gradient accumulation, and mask preprocessing runs once per training step for every method.
{Two workloads are evaluated.
The first uses fixed-length sequences from 1K to 32K with HSTU, RM, CLEAR, and KSA masks, where KSA starts at 2K.
The second uses HSTU masks on real-world sequence lengths from KuaiRand-1K~\cite{gao2022kuairand}.
Besides the two FlexAttention backends, we compare with FlashMask on the masks it can express and with the official KSA kernel on KSA.}}

{{On fixed-length sequences, Figure~\ref{fig:qwen3_hstu_end_to_end} shows that \sys~matches or exceeds the training throughput of every baseline in all configurations.
Across the four masks, \sys~is up to $1.26\times$ faster than Flex-Triton and up to $1.07\times$ faster than Flex-Flash, and on KSA it is $1.01$--$1.07\times$ faster than the official KSA kernel.}
The margins are smaller than in single-operator benchmarks because attention accounts for only part of each training step.}

\begin{table}[t]
\centering
\caption{{End-to-end training throughput on KuaiRand length traces in kTok/s, averaged over three runs.
$^\dagger$FlashMask cannot express packed HSTU masks, so it runs on padded batches.}}
\label{tab:kuairand_trace_e2e}
{\footnotesize
\setlength{\tabcolsep}{5pt}
\begin{tabular}{@{}lrr@{}}
\toprule
Method & Qwen3-8B & Qwen3-14B \\
\midrule
Flex-Triton & $7.59 \pm 0.03$ & $4.49 \pm 0.01$ \\
Flex-Flash & $8.60 \pm 0.06$ & $4.94 \pm 0.01$ \\
FlashMask$^\dagger$ & $4.04 \pm 0.03$ & $2.24 \pm 0.02$ \\
\sys & $\boldsymbol{9.53 \pm 0.09}$ & $\boldsymbol{5.31 \pm 0.03}$ \\
\bottomrule

\end{tabular}}
\vspace{-10pt}
\end{table}

{{On the KuaiRand-1K lengths, Table~\ref{tab:kuairand_trace_e2e} shows} that \sys~achieves the highest throughput for both models, with speedups of $1.11\times$ and $1.08\times$ over Flex-Flash for Qwen3-8B and Qwen3-14B, respectively.
\sys~also outperforms Flex-Triton and FlashMask on this workload.}

\section{Related Work}
\label{sec:related}

\paragraph{Efficient Attention.}
The FlashAttention family~\cite{FlashAttention-NueIPS22,FlashAttention-2_ICLR24,FlashAttention-3_NeurIPS24,FlashAttention-4_MLSys26} improves attention efficiency through IO-aware tiling, work partitioning, and hardware-specific pipelines.
These techniques reduce intermediate memory traffic and improve GPU utilization for long sequences.
FlashInfer~\cite{FlashInfer_MLSys25} extends efficient attention execution to diverse KV-cache layouts and dynamic inference workloads, with customizable attention variants supported through JIT compilation.
Sparse attention methods such as MoBA~\cite{MoBA_Lu_2025}, NSA~\cite{yuan-etal-2025-native}, and DeepSeek Sparse Attention~\cite{DeepSeek-V3.2_DeepSeek-AI_2025} reduce attention computation through content-dependent token or block selection.
Our focus is on efficiently executing user-specified visibility rules.

\paragraph{Attention Customization and Mask Representations.}
FlexAttention~\cite{FlexAttention_MLSys25} exposes score and mask modifications that are compiled into fused attention kernels.
{MetaAttention}~\cite{AttentionEngine_arXiv25} combines programmable attention templates with cross-platform scheduling, while Flashlight~\cite{FlashLight_MLSys26} transforms {attention programs written in native PyTorch} into fused kernels without requiring static templates.
FlashMask~\cite{FlashMask_ICLR25} uses a compact column-wise interval representation for efficient masking, but its fixed four-vector format limits the patterns it can express.
{FlashMask-3~\cite{FlashMask-3_NeurIPS26} brings this representation to Hopper GPUs with a mask-aware kernel and extends it to context-parallel training, while keeping the same four-vector format.}
SPLAT~\cite{SPLAT_OOPSLA_2025} exploits geometric regularity through an affine-compressed sparse-row format and specialized code generation.
SAS~\cite{SAS_EuroSys26} composes static and dynamic sparse patterns and generates inference kernels and KV-cache management functions.
{Similar to the block mask of FlexAttention,} SplashAttention~\cite{SplashAttention_Software24} preprocesses masks into block-level information for sparse attention execution on TPUs.
\sys~compiles user-specified visibility rules into preprocess kernels that instantiate an Execution Descriptor from per-sequence parameters.
The compiler derives the number of intervals from the rules rather than imposing a fixed number, and moves {the evaluation of flip point expressions} out of attention execution.
This separates expressive mask specification from efficient runtime visibility checks.

\paragraph{GPU Programming and Compilation.}
GPU programming systems such as Triton and ThunderKittens~\cite{Triton_MAPL19,ThunderKittens_ICLR25} simplify kernel development through tile-oriented abstractions and support efficient mapping onto GPU hardware.
TileLang~\cite{TileLang_ICLR26} provides a composable tiled programming model and serves as the target for our generated preprocess kernels.
Complementary compiler research explores composable sparse formats and transformations in SparseTIR~\cite{SparseTIR_ASPLOS23}, and joint algebraic and scheduling optimizations in Mirage~\cite{Mirage_OSDI25}.
These systems provide complementary mechanisms for kernel implementation and optimization, while \sys~derives {tile-local execution metadata} from attention visibility rules.

\section{Conclusion}
\label{sec:conclusion}
{
We presented \sys, a compiler-based approach to masked attention that bridges the gap between expressiveness and efficiency by decoupling the predicate users write from the representation the attention kernel executes.
We designed a compiler that lowers a mask predicate into \texttt{MaskSpec}, a Region-based intermediate representation, and generates a mask-specific preprocessing kernel that builds an Execution Descriptor for each batch from the per-sequence parameters.
We further proposed Tile-local Records, which describe the visibility of each row within a tile in a single 32-bit word, or in a bitmap when the row has more than four flip points in the tile, and let the attention kernel check each element with a few integer operations regardless of how many intervals the row has.
Experiments show that \sys~outperforms FlexAttention and matches or exceeds FlashMask in single-operator benchmarks across a range of masks and improves end-to-end training throughput, while users write only the predicate.}

\bibliography{CompileMask}
\bibliographystyle{compilemask}

\clearpage
{\centering{\LARGE\bf Appendix}\par\vskip 0.25in}
\appendix
\section{{Mask Shapes}}
\label{sec:app_mask_shapes}

\begin{figure*}[t]
\centering
\includegraphics[width=\textwidth]{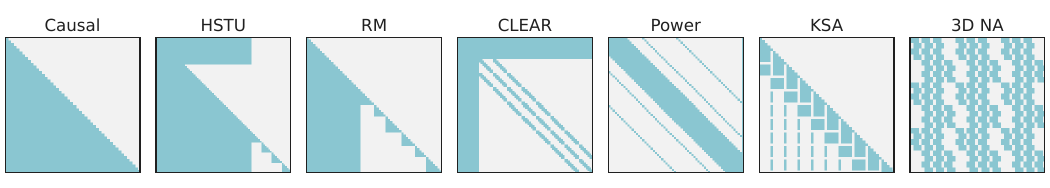}
\vspace{-4pt} %
\caption{{Shapes of the evaluated masks on small sequences, where rows are queries, columns are keys, and teal marks visible elements.
The sizes are scaled down for readability.
HSTU uses 80 tokens with a 16-token context, a 40-token history, and 24 targets in groups of 6, so the targets are enlarged relative to the experiments to make the groups visible.
RM uses a 24-token prompt and six responses, and CLEAR uses 12 text tokens and an $8\times8$ image grid with $R=2$.
KSA uses chunks of four text tokens and one summary token with two preceding chunks, and is shown in its logical interleaved order without the summary copies.
3D NA uses a $3\times4\times4$ grid with a $3\times3\times3$ window.}}
\label{fig:app_mask_shapes}
\end{figure*}

{Figure~\ref{fig:app_mask_shapes} illustrates the shape of each mask evaluated in Section~\ref{sec:experiment} on a small sequence.
HSTU splits each sequence into a context, a history, and targets in groups, where context queries see the context and history, history queries attend causally, and target queries see the complete context and history together with the causal part of their own group.
The reward-model mask places six responses after a shared prompt, and each response attends to the prompt and causally to itself.
CLEAR concatenates a text prefix with an image grid, where text queries attend globally and image queries attend to all text tokens and to the image tokens within a spatial radius $R$.
Power lets each query attend to its local neighborhood, to every key at a power-of-two distance, and to the last key.
KSA lets each text token attend to the text of its own chunk and of the recent chunks and to the summary tokens of older chunks.
3D NA lets each video token attend to a three-dimensional window around it.}

\begin{figure}[t]
\centering
\includegraphics[width=\columnwidth]{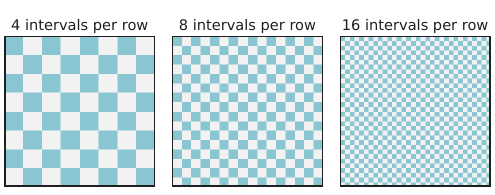}
\caption{{Checkerboard masks with 4, 8, and 16 visible intervals per row on a 128-token sequence, where teal marks visible elements.}
{The experiments use about 4K tokens and from 1 to 256 visible intervals per row.}}
\label{fig:app_checkerboard_shapes}
\end{figure}

{Figure~\ref{fig:app_checkerboard_shapes} illustrates the checkerboard masks used in Section~\ref{subsec:records_vs_intervals}, in which visible and masked square cells of equal size alternate along both axes, starting from a visible cell at the top left, without a causal constraint.
Each query row therefore alternates between visible and masked intervals of equal length, and rows in adjacent bands of cells start from opposite states.}

\section{{Experimental Settings}}
\label{sec:app_settings}

{All experiments run on NVIDIA Hopper GPUs with BF16 precision, PyTorch 2.13.0, and CUDA 13.0.
\sys~uses TileLang 0.1.9 for preprocess kernels and CUTLASS DSL 4.8.0.dev0 for attention kernels.
Flex-Flash uses {the SM90 kernels of FlashAttention-4~\cite{FlashAttention-4_MLSys26}, implemented in CuTeDSL,} at commit 47e91f1.
The attention kernels of \sys~are derived from these kernels and maintained as a separate implementation.
In single-operator benchmarks, FlashMask uses its official CuTeDSL implementation at commit b492328 with PaddlePaddle 3.3.0.
End-to-end training uses a PyTorch port of the same FlashMask kernels.
For single-operator benchmarks, we lock the GPU clock, flush the L2 cache before each measurement, and average repeated measurements after warmup.
End-to-end training measures 50 steps after 10 warmup steps and averages the throughput over three runs.
Table~\ref{tab:app_settings_summary} summarizes the workload parameters of each mask, and Table~\ref{tab:app_na_configs} lists the configurations of 3D NA.}

\begin{table*}[!t]
\centering
\caption{{Workload parameters of each mask in the single-operator benchmarks, fixed-length end-to-end training, and KuaiRand end-to-end training.
A dash marks an experiment that does not evaluate the mask.}}
\label{tab:app_settings_summary}
\footnotesize
\begin{tabular}{@{}lp{0.3\textwidth}p{0.27\textwidth}p{0.22\textwidth}@{}}
\toprule
Mask & Single-operator & Fixed-length end-to-end & KuaiRand \\
\midrule
HSTU & Context, history, and targets at about 20\%, 70\%, and 10\%, $g$ uniform in $[3,8]$ per sequence & Same allocation, $g=8$ sampled once with a fixed seed & {Context of 30 and eight targets with $g=1$, history from user-day traces, eight users per step} \\
Reward model & Six responses, each of length in $[\lfloor N/16\rfloor,\lfloor N/11\rfloor]$, prompt fills the rest & Same distribution, sampled once with a fixed seed & -- \\
CLEAR & $R=2$, text lengths 185, 140, 268, and 253 for the four instances, image grids from $32\times32$ to $256\times256$ & $R=2$, text length varies per step in $[128,384]$, image grids from $32\times32$ to $256\times256$ & -- \\
Causal & Sequence length only & {--} & -- \\
Power & Sequence length only & -- & -- \\
KSA & Eight text tokens and one summary token per chunk, 128 recent chunks & {Same layout, text lengths from 2K to 32K} & -- \\
3D NA & Four configurations in Table~\ref{tab:app_na_configs} & -- & -- \\
\bottomrule
\end{tabular}
\end{table*}

\begin{table}[t]
\centering
\caption{{Configurations of the 3D NA workload.
All configurations use 24 attention heads with a head dimension of 128.}}
\label{tab:app_na_configs}
\footnotesize
\resizebox{\columnwidth}{!}{%
\begin{tabular}{@{}lcccc@{}}
\toprule
Configuration & Grid $T\times H\times W$ & Window & Tokens & Batch \\
\midrule
STA & $30\times48\times80$ & $19\times25\times25$ & 115,200 & 1 \\
Hunyuan & $30\times48\times80$ & $30\times40\times40$ & 115,200 & 1 \\
Cosmos-low & $16\times44\times80$ & $16\times32\times48$ & 56,320 & 2 \\
Cosmos-high & $16\times44\times80$ & $16\times24\times16$ & 56,320 & 2 \\
\bottomrule
\end{tabular}}
\end{table}

\section{Numerical Correctness}
\label{sec:app_correctness}

{We compare \sys~against dense masked attention on an NVIDIA Hopper GPU, using the same attention kernels with Tile-local Records as the single-operator experiments in Section~\ref{sec:experiment}.
Queries, keys, and values are drawn in BF16 from $\mathcal{N}(0,0.5^2)$ with a fixed seed.
The reference casts the same inputs to FP32 and computes}
\begin{equation}
\begin{aligned}
P_{\mathrm{ref}}&=\operatorname{softmax}\!\left(QK^{\top}/\sqrt{d}+M\right),\\
O_{\mathrm{ref}}&=P_{\mathrm{ref}}V,
\end{aligned}
\end{equation}
{where $M$ is zero at visible positions and $-\infty$ elsewhere.
The reference evaluates the visibility rule of each mask directly rather than through the compiler, and its matrix products run in FP32 without TF32.
Backward checks use the same upstream gradient for \sys~and the reference, and compare $dQ$, $dK$, and $dV$ obtained through differentiation.}
For each tensor $X$, we report the maximum absolute error
\begin{equation}
E_{\max}(X)=\max_t\left|X_t-X_{\mathrm{ref},t}\right|.
\end{equation}
{A check passes when every element satisfies}
\begin{equation}
\left|X_t-X_{\mathrm{ref},t}\right|
\le \epsilon_{\mathrm{abs}}+\epsilon_{\mathrm{rel}}\left|X_{\mathrm{ref},t}\right|,
\end{equation}
{where both tolerances are $0.04$ for outputs and $0.08$ for gradients.}

\begin{table}[t]
\centering
\caption{{Maximum absolute error of \sys~against FP32 dense masked attention.
Lengths count image tokens for CLEAR and text tokens for KSA, and 3D NA uses the STA configuration.
All 12 configurations pass.}}
\label{tab:app_correctness}
\footnotesize
\setlength{\tabcolsep}{3pt}
\begin{tabular*}{\columnwidth}{@{\extracolsep{\fill}}llrrrr@{}}
\toprule
Mask & Length & $O$ & $dQ$ & $dK$ & $dV$\\
\midrule
HSTU & 128 & 1.22e-03 & 7.16e-04 & 6.75e-04 & 2.45e-03 \\
HSTU & 1K & 4.61e-04 & 2.59e-04 & 2.81e-04 & 1.10e-03 \\
CLEAR & 1K & 6.93e-04 & 3.72e-04 & 7.60e-04 & 2.96e-03 \\
CLEAR & 2K & 6.97e-04 & 3.82e-04 & 1.12e-03 & 4.85e-03 \\
CLEAR & 4K & 7.07e-04 & 3.79e-04 & 1.41e-03 & 5.64e-03 \\
CLEAR & 8K & 7.94e-04 & 4.24e-04 & 2.44e-03 & 9.17e-03 \\
CLEAR & 16K & 7.81e-04 & 3.95e-04 & 2.71e-03 & 1.15e-02 \\
Power & 1K & 1.48e-03 & 1.21e-03 & 3.18e-03 & 1.41e-02 \\
Power & 4K & 1.50e-03 & 1.40e-03 & 5.72e-03 & 1.88e-02 \\
KSA & 1K & 3.49e-03 & 2.72e-03 & 2.98e-03 & 1.20e-02 \\
KSA & 4K & 4.07e-03 & 3.18e-03 & 2.06e-03 & 1.54e-02 \\
3D NA & STA & 8.93e-05 & 4.96e-05 & 7.73e-05 & 3.05e-04 \\
\bottomrule

\end{tabular*}
\end{table}

{Table~\ref{tab:app_correctness} covers five masks in 12 configurations.
HSTU is checked at lengths of 128 and 1,024 with a batch size of two.
CLEAR is checked at image lengths from 1K to 16K with 185 text tokens, which matches the first CLEAR instance in Figure~\ref{fig:operator_fwd_bwd}, and with the batch sizes of the single-operator experiments.
Power and KSA are checked at lengths of 1K and 4K with a batch size of two.
3D NA is checked on the full-size STA configuration in Table~\ref{tab:app_na_configs} with 24 heads.
The CLEAR, Power, and KSA configurations include row segments that use the bitmap fallback.
Across all configurations, the largest output error is $4.07\times10^{-3}$ and the largest gradient error is $1.88\times10^{-2}$, both well below the tolerances.}

\section{{Record Decoding and Cost Analysis}}
\label{sec:app_records_theory}

{This appendix gives the main loop of the attention kernel, proves Theorem~\mainref{thm:record_decoding}, and derives the costs in Table~\mainref{tab:records_cost}.
{Throughout, $N_q$ and $N_k$ denote the numbers of queries and keys, and a tile covers $B_r$ query rows and $B_c$ key columns of the score matrix.
A forward Record spans the $B_c=128$ key columns of a tile and a backward Record spans the 64 query rows of a tile, so the span of a Record never exceeds 128 positions.
We write $c\in[0,128)$ for a position within the span of a Record.}
All arithmetic on Records is 32-bit unsigned arithmetic, that is, modulo $2^{32}$.}

\subsection{{Attention Kernel Loop}}
\label{subsec:app_kernel_loop}

\begin{algorithm}[t]
\caption{{Attention execution for one query block.}}
\label{alg:records_attention}
\begin{algorithmic}[1]
\REQUIRE query block $Q$, key and value tiles $K_n$ and $V_n$, Tile-State Bitmaps, Row States, and Tile-local Records
\STATE $r_i \gets \textsc{FetchNext}(i)$ for each row $i$
\FOR{each key tile $n$ with $\mathrm{compute}[n]=1$}
  \STATE $S \gets Q K_n^{\top}$
  \IF{$\mathrm{mask}[n]=1$}
    \FOR{each row $i$}
      \STATE $S_{i,c} \gets -\infty$ where $\textsc{Decode}(r_i, c)=0$
      \STATE $r_i \gets \textsc{FetchNext}(i)$
    \ENDFOR
  \ENDIF
  \STATE online softmax on $S$, then $O \gets O + P V_n$
\ENDFOR
\end{algorithmic}
{\small $\textsc{FetchNext}(i)$ reads the next state of row $i$ and returns the next Record of the row if the state is 10, or the constant Record of state 01 or 00 otherwise.
$\textsc{Decode}$ is the decoder of Theorem~\mainref{thm:record_decoding}.}
\end{algorithm}

{Algorithm~\ref{alg:records_attention} gives the main loop of the attention kernel for one query block in the forward pass.
The kernel scans each 32-bit word of the compute bitmap with bit-scan instructions, so it never visits an empty tile.
The mask bit of a tile is the same for all threads of a thread block, so the branch in line~4 causes no divergence.
Full and partial tiles run two separately compiled versions of the loop body, and only the version for partial tiles contains masking code.
Consecutive full tiles are processed together in an inner loop.
Line~7 issues the load of the next Record right after the mask of the current tile is applied, so the load overlaps with the rest of the softmax of the current tile and with the $QK^\top$ of the next tile.
The backward pass uses the same loop over the query tiles of a key block.}

\subsection{{Proof of Theorem~\mainref{thm:record_decoding}}}

{Consider a mixed row segment with initial visibility $\mathrm{init}\in\{0,1\}$ and flip points $1\le p_1<\dots<p_m\le 127$, where $m\le 4$.
Its Record is
\[
w = \mathrm{init}\cdot 2^{31} + \sum_{k=1}^{m} (128-p_k)\cdot 2^{8(k-1)},
\]
and the bytes of unused fields are zero.
Byte $k-1$ of $w$ therefore holds $128-p_k\le 127$ for $k<4$, so bits 7, 15, and 23 of a Record are zero, and byte 3 holds $\mathrm{init}\cdot 128 + (128-p_4)$ when $m=4$.}

{Equation~\mainref{eq:flip_parity} holds by the definition of a flip point.
Moving from column 0 to column $c$, the visibility of the row changes exactly once at every flip point $p_k\le c$, so $\operatorname{visible}(c)$ equals $\mathrm{init}$ flipped once for every such $p_k$.
It remains to show that \textsc{Decode} in Equation~\mainref{eq:decode} computes this parity.}

{Let $u=\mathtt{0x01010101}$ and $x=w+c\,u$, which adds $c$ to every byte of $w$.
For bytes 0, 1, and 2, a used field becomes $(128-p_k)+c\le 127+127=254$, so the addition never carries into the next byte.
Its top bit is one exactly when $(128-p_k)+c\ge 128$, that is, when $c\ge p_k$.
An unused field becomes $c\le 127$, so its top bit is zero.
For byte 3, any carry out of bit 31 leaves the 32-bit word, so bit 31 of $x$ equals $\mathrm{init}\oplus[c\ge p_4]$ when $m=4$ and equals $\mathrm{init}$ otherwise.}

{Let $a_0,\dots,a_3$ be the top bits of the four bytes of $x$, so that $y = x \mathbin{\&} 128u = a_0 2^{7} + a_1 2^{15} + a_2 2^{23} + a_3 2^{31}$.
The product $y\cdot u = y\,(1+2^{8}+2^{16}+2^{24})$ contains the four terms $a_0 2^{31}$, $a_1 2^{31}$, $a_2 2^{31}$, and $a_3 2^{31}$, whose sum is $(a_0+a_1+a_2+a_3)\,2^{31}$.
Every other term below $2^{31}$ belongs to the sum $a_0(2^{7}+2^{15}+2^{23}) + a_1(2^{15}+2^{23}) + a_2 2^{23}$, which is less than $2^{31}$ and therefore never carries into bit 31.
All remaining terms are multiples of $2^{32}$ and vanish.
Bit 31 of $y\cdot u$ is therefore $(a_0+a_1+a_2+a_3)\bmod 2$, and the final right shift by 31 returns this bit, which equals $\mathrm{init}\oplus\bigoplus_{k=1}^{m}[c\ge p_k]$.}

{Because exclusive or is commutative, the result does not depend on which field holds which flip point.
For the constant Record $\mathtt{0x80000000}$, every byte of $x$ other than byte 3 equals $c\le 127$, and byte 3 equals $128+c$ modulo 256, so the result is 1 for every column.
For the constant Record $0$, every byte equals $c\le 127$, so the result is 0 for every column.
This completes the proof.}

{The proof requires that $c$ and $128-p_k$ each fit in seven bits, which holds because {the span of a Record never exceeds 128 positions}.
A Record whose bit 7 is set marks the bitmap fallback for a segment with more than four flip points.
The attention kernel reads the exact visibility bitmap of that segment instead of decoding the Record, so the mask stays exact in this case as well.}

\subsection{{Cost of Masking in the Attention Kernel}}

{Consider query row $i$, and let $f_i$ be its number of flip points over the whole row.
A row with at most $K$ intervals has $f_i\le 2K$.
Let $P_i$ be the number of partial tiles in the query block of row $i$, which are the tiles of this block that contain both visible and masked elements.
Let $M_i$ be the number of mixed segments of row $i$, and $F_i$ the number of its segments that use the bitmap fallback.}

{\textbf{Operations per element.}
For non-fallback Records, by Theorem~\mainref{thm:record_decoding}, the visibility of an element takes one multiplication, one addition, one bitwise and, one more multiplication, and one shift.
For a fallback segment, the bitmap lookup is also $O(1)$.
This count does not depend on $K$ or $f_i$.
With per-row intervals, the kernel compares each element with all $K$ intervals of its row and combines the results, which takes $O(K)$ operations.}

{\textbf{Registers per row.}
A row keeps its current Record, the prefetched Record of its next partial tile, and a position in its sequence of states, which is $O(1)$ registers.
With per-row intervals, a row keeps the two endpoints of each of its $K$ intervals, which is $O(K)$ registers.}

{\textbf{Descriptor loads.}
The kernel reads exactly the words that the descriptor stores for row $i$, so the following bounds cover both storage and reads.}

{\textit{Intervals or Records.}
Row $i$ uses one Record word for each mixed segment and up to four additional words for each segment that uses the bitmap fallback, which gives at most $M_i+4F_i$ words.
A mixed segment that does not use the fallback contains at least one flip point and takes one word.
A fallback segment contains more than four flip points and takes at most five words.
Every segment therefore takes at most one word per flip point that it contains, so row $i$ uses at most $f_i\le 2K$ Record words, which is $O(K)$.
With per-row intervals, the descriptor stores the $K$ intervals of a row and each CTA loads them once, which is also $O(K)$ words.}

{\textit{Row States.}
For every partial tile, row $i$ has a two-bit state, and sixteen states fit in one word, which gives $\lceil P_i/16\rceil$ words, that is, $O(P_i/16)$.
Since a row passes at most $\lceil N_k/B_c\rceil$ tiles, this amounts to about one word for every $16B_c$ key columns.
Per-row intervals need no Row States.
The Tile-State Bitmaps add two bits per tile, which is $O(N_q N_k/(B_r B_c))$ bits for the whole mask.}

\subsection{{Cost of Preprocessing}}

{We analyze the case in which the visible columns of every Region form a union of intervals whose flip point expressions can be evaluated directly.
The preprocess kernel assigns one thread to each query row.
The thread evaluates the $O(K)$ flip points of its row and sorts them.
It then visits only the partial tiles of its query block.
For each such tile, it writes the visibility of its row into $B_c/32$ words, extracts the flip points from these words with bitwise operations, and emits the state and the Record of the row.
Fully visible tiles are marked with bitwise operations, and empty tiles are skipped.
After sorting, encoding a row within partial tiles takes $O(P_i\,B_c/32 + f_i)$ work.
This bound excludes the initialization and construction of the Tile-State Bitmaps.
The Tile-State Bitmaps are obtained by reducing the per-row visibility across the rows of each tile, and the backward descriptor is obtained in the same pass by transposing the visibility of each tile.}

\end{document}